\documentstyle[emulateapj] {article}

\lefthead{EVANS ET AL.}
\righthead{CO IN PKS 1345+12}

\begin{document}

\title{ Molecular Gas in the Powerful Radio Nucleus of the Ultraluminous
Infrared Galaxy PKS 1345+12}

\author{A. S. Evans\altaffilmark{1,2},
D. C. Kim\altaffilmark{3,4},
J. M. Mazzarella\altaffilmark{3},
N. Z. Scoville\altaffilmark{1},
\and
D. B. Sanders\altaffilmark{6}}
 
\altaffiltext{1}{Division of Physics, Math, \& Astronomy,
California Institute of Technology, Pasadena, CA 91125}
\altaffiltext{2}{Email Address: ase@astro.caltech.edu}
\altaffiltext{3}{Infrared Processiong \& Analysis Center, California
Institute of Technology, MS 100-22, Pasadena, CA 91125}
\altaffiltext{4}{Present Address: Academia Sinica Institute of Astronomy \&
Astrophysics P.O. Box 1-87, NanKang Taipei 115, Taiwan, R. O. C.}
\altaffiltext{5}{UCO/Lick Observatory, University of California,
Santa Cruz, CA 95064}
\altaffiltext{6}{Institute for Astronomy, 2680 Woodlawn Drive, Honolulu,
HI 96822}

\begin{abstract}

Millimeter CO($1\to0$) interferometry and high resolution, {\it Hubble
Space Telescope} (HST) 1.1, 1.6, and 2.2 $\mu$m imaging of the radio
compact galaxy PKS 1345+12 are presented. With an infrared luminosity of
$\sim 2\times10^{12}$ L$_\odot$, PKS 1345+12 is a prime candidate for
studying the link between the ultraluminous infrared galaxy phenomenon and
radio galaxies.  These new observations probe the molecular gas
distribution and obscured nuclear regions of PKS 1345+12 and provide
morphological support for the idea that the radio activity in powerful
radio galaxies is triggered by the merger of gas rich galaxies.  Two
nuclei separated by 2$\arcsec$ (4.0 kpc) are observed in the
near-infrared; the extended southeastern nucleus has colors consistent
with reddened starlight,  and the compact northwestern nucleus has
extremely red colors indicative of an optical quasar with a warm dust
component.  Further, the molecular gas, 3mm continuum, and radio emission
are coincident with the redder nucleus, confirming that the northwestern
nucleus is the site of the AGN and that the molecular gas is the likely
fuel source.

\end{abstract}


\section{Introduction}

Almost all ultraluminous infrared galaxies [ULIGs: $L_{\rm IR} (8-1000
\mu$m) $\ge 1\times10^{12}$ L$_\odot$] have optical/near-infrared
morphologies indicative of galaxy-galaxy mergers (e.g. Joseph \& Wright
1985; Armus, Heckman, \& Miley 1987; Sanders et al. 1988a; Murphy et al.
1996; Kim 1995).  Their large quantities of dust and molecular gas (e.g.,
Sanders, Scoville, \& Soifer 1991; Solomon et al.  1997),  as well as
evidence of abundant young star clusters in many ULIGs (e.g.  Surace et
al. 1998), are strong evidence of recent/ongoing star formation.
Molecular gas is also a likely source of fuel for active galactic nuclei
(AGN) such as quasars and radio galaxies, many of which have high $L_{\rm
IR}$ and  disturbed optical morphologies (e.g., Stockton \& MacKenty 1983;
MacKenty \& Stockton 1984; Heckman et al. 1986; Smith \& Heckman 1989a,b),
thus providing a possible connection between mergers and the building of
supermassive nuclear black holes.

PKS 1345+12 (IRAS 13451+1232: $L_{\rm IR} = 1.7\times10^{12}$ L$_\odot$)
is a prime candidate for the link between the ULIG phenomenon and radio
galaxies. With a radio luminosity of $P_{\rm 408 MHz} = 2.4\times10^{26}$
W Hz$^{-1}$, it is the most powerful radio galaxy detected in CO($1\to0$)
to date.  It also belongs to a family of ``warm'' ($f_{\rm 25\mu m}/f_{\rm
60\mu m} \ge 0.2$, similar to the colors of Seyfert galaxies:  de Grijp,
Miley, \& Lub 1987) infrared galaxies believed to be in a transition state
between the ``cold'' ($f_{\rm 25\mu m}/f_{\rm 60\mu m}< 0.2$) ULIG
phenomenon, when rampant star-formation is occuring and the accretion disk
is forming around the nuclear black hole, and the optical quasar phase
(Sanders et al 1988a,b).  It is observed to have two nuclei with a
projected separation of $\sim 2\arcsec$ ($\sim$4 kpc: Heckman et al.
1986; Smith \& Heckman 1989a,b; Kim 1995), a very compact radio jet
($0.1\arcsec \sim 200$ pc), and an extremely high molecular gas mass
($4.4\times10^{10}$ M$_\odot$:\footnote{Assuming an $L^{\prime}_{\rm CO}$
to $M_{\rm H_2}$ conversion factor of 4 M$_\odot$ (K km s$^{-1}$
pc$^2$)$^{-1}$; see \S 3.} Mirabel, Sanders, \& Kazes 1989).  The ratios
of the narrow optical emission lines in PKS 1345+12 indicate that it
contains a Seyfert 2 nucleus (Sanders et al.  1988b; Veilleux et al.
1995).  Further, recent near-infrared spectroscopic observations have
detected broad ($\Delta{\rm v_{FWHM}} \sim$ 2600 km s$^{-1}$) Pa$\alpha$
emission, indicating the presence of a quasar nucleus which is obscured at
optical wavelengths (Veilleux, Sanders, \& Kim 1997).

The near-infrared spectroscopy of PKS 1345+12 illustrates the importance
of probing the nature and distribution of obscured nuclear energy sources
in ULIGs at near-infrared wavelengths. Such studies also complement CO
interferometry, which provides spatial and kinematic information of the
molecular gas reservoirs in these dusty systems. In this {\it Letter}, the
superior resolution of the HST\footnote{The NASA/ESA Hubble Space
Telescope is operated by the Space Telescope Science Institute managed by
the Association of Universities for Research in Astronomy Inc.  under NASA
contract NAS5-26555.} Near-Infrared Camera and Multi-Object Spectrometer
(NICMOS:  0.1--0.2$\arcsec$ resolution at 1--2 $\mu$m) and the Owens
Valley Millimeter Array\footnote{The Owens Valley Millimeter Array is a
radio telescope facility operated by the California Institute of
Technology and is supported by NSF grants AST 93--14079 and AST
96--13717.} ($\sim 2\arcsec$ resolution at 3 mm) are used to show for the
first time the spatial coincidence of the molecular gas and the radio-loud
nucleus of PKS 1345+12. These observations provide further support for the
relationship between mergers, molecular gas, and AGN activity.  An $H_0 =
75$ km s$^{-1}$ Mpc$^{-1}$ and $q_0= 0.0$ are assumed throughout such that
1$\arcsec$ subtends $\sim$ 2.0 kpc at the redshift of the galaxy
($z=0.1224$).

\section{Observations and Data Reduction}

\subsection{NICMOS Observations}

HST  NICMOS observations of PKS 1345+12 were obtained as part of a larger
program to image infrared-luminous galaxies (Scoville
et al.  1999; Evans 1999a). Observations were obtained in a single orbit
on 1997 December 5 using camera 2, which consists of a $256\times256$
array with pixel scales of 0.0762$\arcsec$ and 0.0755$\arcsec$ per pixel
in $x$ and $y$, respectively, providing a $\sim 19.5\arcsec
\times19.3\arcsec$ field of view (Thompson et al. 1998).  Images were
obtained using the wide-band filters F110W (1.10$ \mu$m, $\Delta
\lambda_{\rm FWHM} \sim 0.6 \mu$m), F160W (1.60 $\mu$m, $\Delta
\lambda_{\rm FWHM} \sim 0.4 \mu$m), and F222M (2.22 $\mu$m, $\Delta
\lambda_{\rm FWHM} \sim 0.14 \mu$m), which provide a 
resolution (FWHM) of 0.11$\arcsec$, 0.16$\arcsec$, and
0.22$\arcsec$, respectively. The basic observation and data reduction
procedures are the same as those described in Scoville et al. (1999); the
total integration times per filter setting for these observations were 480
sec (1.1 and 1.6 $\mu$m) and 600 sec (2.2 $\mu$m).  Flux calibration of
the images were done using the scaling factors $2.03\times10^{-6},
2.19\times10^{-6}$, and $5.49\times10^{-6}$ Jy (ADU/sec)$^{-1}$ at 1.10,
1.60, and 2.22 $\mu$m, respectively. The corresponding magnitudes were
calculated using the zeropoints 1775, 1083, and 668 Jy (Rieke 1999).

\subsection{Interferometric Observations}

Aperture synthesis maps of CO($1\to0$) and 2.7 mm continuum emission in
PKS 1345+12 were made with the Owens Valley Radio Observatory (OVRO)
Millimeter Array during five observing periods from 1996 September to 1997
May. The array consists of six 10.4 m telescopes, and the longest observed
baseline was 242 m. Each telescope was configured with $120\times4$ MHz
digital correlators. During the observations, the nearby quasar HB89
1413+135 (1.5 Jy at 103 GHz; 14$^{\rm h}$13$^{\rm m}$33.92$^{\rm s}$
+13$^{\rm o}$:34$\arcmin$17.51$\arcsec$ [B1950.0]) was observed every 25
minutes to monitor phase and gain variations, and 3C 273 and 3C 454.3 were
observed to determine the passband structure. Finally, flux calibration
observations of Uranus were obtained.

The OVRO data were reduced and calibrated using the standard Owens Valley
data reduction package MMA (Scoville et al. 1992). The data were then
exported to the mapping program DIFMAP (Shepherd, Pearson, \& Taylor
1995).

\section{Results}

The reduced 1.1, 1.6, and 2.2 $\mu$m images are shown in Figure 1a--c. The
galaxy consists of two nuclei with a projected separation of $2.0\arcsec$
(4 kpc). Low level surface brightness emission envelops both nuclei, with
an east-west extent of 11$\arcsec$ (22 kpc; full width at 0.5\% the
maximum flux density at 1.1 and 1.6 $\mu$m), and a 5.5$\arcsec$ (11 kpc)
southern extent beyond the southeastern nucleus (hereafter PKS
1345+12SE).  The radial surface brightness profile does not constrain the
nature of the progenitor galaxies or the type of galaxy they are evolving
into; both an $r^{0.25}$ law and an exponential disk give reasonable fits
to the profile (see also Scoville et al.  1999).

PKS 1345+12SE is observed to be extended at all three wavelengths with a
FWHM of 0.15$\arcsec$ (300 pc) at 1.1 $\mu$m. The measured 1.1$\arcsec$
aperture magnitudes are 16.86, 15.85, and 15.31 at 1.1, 1.6, and 2.2
$\mu$m, respectively, and the derived colors are thus $m_{1.1-1.6}$ =1.01
and $m_{1.6-2.2}$ = 0.54. In contrast, the northwestern nucleus (hereafter
PKS 1345+12NW) is unresolved with a FWHM of 0.11$\arcsec$ (220 pc) at 1.1
$\mu$m and magnitudes of 16.66, 15.45, and 13.96 at at 1.1, 1.6, and 2.2
$\mu$m, respectively. The near-infrared colors of PKS 1345+12NW are
extremely red; $m_{1.1-1.6}$ =1.20 and $m_{1.6-2.2}$ = 1.49.  The red
nature of PKS 1345+12NW relative to PKS 1345+12SE is also evident in the
1.1$\arcsec$-aperture flux density ratios of the two nuclei;  $f({\rm
SE})/f({\rm NW})$ decreases from a value of 0.82 at 1.1 $\mu$m to 0.29 at
2.2 $\mu$m.

Both the CO($1\to0$) emission and underlying 2.7 mm continuum in PKS
1345+12 are unresolved.  The continuum flux density is 0.31 Jy and is
consistent with a power-law extrapolation of the radio flux density (e.g.
Steppe et al. 1995). The CO emission (Figure 2) has a $\Delta{\rm v}_{\rm
FWHM} \sim600$ km s$^{-1}$, a flux density of 14$\pm4$ Jy km s$^{-1}$, and
a CO luminosity of $L^{\prime}_{\rm CO} = 8.2\times10^{9}$ K km
s$^{-1}$ pc$^2$.  Thus, the line profile and luminosity are consistent
with those derived from NRAO 12m Telescope observations of PKS 1345+12
(i.e., $L^{\prime}_{\rm CO} = 1.1\times10^{10}$ K km s$^{-1}$ pc$^2$:
Mirabel et al. 1989), and confirms that the flux measured in the
single-dish observations (i.e., 81$\arcsec$ FWHM beam) is entirely
recovered with OVRO ($\sim 2.2\arcsec$ synthesized beam).  Assuming a
standard ratio ($\alpha$) of CO luminosity to H$_2$ mass of 4 M$_\odot$ (K
km s$^{-1}$ pc$^2$)$^{-1}$, which is similar to the value determined for
the bulk of the molecular gas in the disk of the Milky Way (Scoville \&
Sanders 1987; Strong et al.  1988), the molecular gas mass is calculated
to be $3.3\times10^{10}$ M$_\odot$, or 14 times the molecular gas mass of
the Milky Way.\footnote{Radford, Solomon, \& Downes (1991) have used
theoretical models to determine that $\alpha$ ranges from 2--5 M$_\odot$
for a reasonable range of temperatures and densities, thus the molecular
gas mass of PKS 1345+12 may actually be as low as 1.6$\times10^{10}$
M$_\odot$.} Finally, using the FWHM beam of the CO map (2.2$\arcsec$), the
molecular gas concentration is calculated to be $>$2000 M$_\odot$
pc$^{-2}$, or $>$15--1500 times that observed in local early-type spiral
galaxies (e.g., Young \& Scoville 1991) and comparable to the
concentrations observed in a sample of luminous infrared galaxies observed
by Scoville et al. (1991) and Bryant \& Scoville (1999).

\section{Astrometry of the Near-Infrared Images}

The interpretation of the data presented in this {\it Letter} depends on
accurate astrometry of the multiwavelength images. To determine the
coordinates of the two nuclei of PKS 1345+12, the positions of stars
within 3.5$\arcmin$ of the galaxy were first retrieved from the USNO-A1.0
database.  A plate solution (world coordinate system) was then derived for
a $7\arcmin\times7\arcmin$ {\it R}-band image (Kim 1995), which also shows
the two nuclei of PKS 1345+12, using the IRAF task PLTSOL.

The coordinates of the two nuclei, along with the positions of the CO and
radio emission from the galaxy, are listed in Table 1. The radio and CO
emission appear to be spatially coincident with PKS 1345+12NW - the
measured near-infrared peak of PKS 1345+12NW is displaced 0.3$\arcsec$ NE
of the radio emission, but is within the uncertainties associated with the
positions of the stars used to derive the position of the PKS 1345+12
nuclei. Likewise, the measured near-infrared peak of PKS 1345+12NW is
displaced $0.44\arcsec$ NE of the CO emission and continuum centroids,
consistent with the measured OVRO beamsize (see Figure 1d).

\section{Discussion}

The bulk of the activity in PKS 1345+12 is associated with PKS 1345+12NW.
The derived NICMOS colors of the two nuclei provides further support that
the AGN resides in PKS 1345+12NW.  While the colors of PKS 1345+12SE are
consistent with starlight reddened by 1--5 magnitudes of dust, PKS
1345+12NW has colors similar to other warm ULIGs observed with NICMOS -
i.e., similar to optically-selected quasars with a 500--1000 K dust
component (see also Scoville et al. 1999 and the summary in Evans 1999a).
Similar results are derived from near-infrared ground-based observations
of PKS 1345+12 (e.g., Surace \& Sanders 1999), and recent, high-resolution
near-infrared spectroscopy (Veilleux \& Sanders 1999) confirm that PKS
1345+12 is the source of the broad-lines detected by Veilleux, Sanders, \&
Kim (1997).

The presence of a large and concentrated reservoir of molecular gas in PKS
1345+12NW is consistent with the notion that this gas is a
source of fuel for the radio phenomenon (e.g., Mirabel et al. 1989). In
this scenario, the molecular gas in the western galaxy is driven inward
via gravitational instabilities induced by interactions with PKS
1345+12SE.

Of all of the radio galaxies detected in CO($1\to0$) to date (Phillips et
al. 1987; Mirabel et al. 1989; Mazzarella et al. 1993; Evans 1999b; Evans
et al.  1999), PKS 1345+12 is the most molecular gas-rich, and the only
one that clearly has two nuclei.  Thus, while PKS 1345+12 is an advanced
merger in terms of its relatively small nuclear separation, the fact that
the stellar nuclei are 4 kpc apart implies that this system is dynamically
younger than the other radio galaxies observed. If the assumption is made
that the nuclei of PKS1345+12 are gravitationally bound, the relative
velocity of the merging galaxies is $|v| \lesssim (GM_{\rm gal}/R_{\rm
sep})^{1/2} \lesssim 300$ km s$^{-1}$, where $M_{\rm gal} = 10^{11}$
M$_\odot$ and $R_{\rm sep} \sim 4$ kpc, and thus the merger has at least
an additional $\sim10^7$ years before the nuclei coalesce.

{}From the extent of the radio emission, it is also clear that the
jet activity commenced fairly recently. In the 2.3 and 8.5 GHz radio maps
shown in Fey, Clegg, \& Fomalont (1996), the radio jet has a maximum
extent of 0.10$\arcsec$ ($\sim 200$ pc).  Thus, if the jet propagates at a
speed of 0.1$c$, it can be no more than 7000 years old. For comparison,
the linear extent of the jets associated with single-nuclei radio galaxies
detected in CO are 10--200 kpc, but their estimated jet ages ($\lesssim
10^7$ years) are significantly less than the timescale of the merger
process ($10^9$ years). The creation of the radio jets so late in the life
of the merger can be understood in terms of merger dynamics - there will
be a natural offset in the time at which the merger begins and the AGN
activity occurs because of the time it takes for the molecular gas to
agglomerate in the nuclear regions of the galaxy.  Further, depending on
how typical such a delay is in radio galaxies, the consumption of
molecular gas by extended star formation may be well underway prior to the
onset of the radio activity, and may continue for another $10^7$ years or
so. This provides a natural explanation of why radio galaxies with older,
extended jets are not observed to have large reservoirs of molecular gas
(i.e., $M({\rm H}_2) \lesssim 10^9$ M$_\odot$:  Mazzarella et al. 1993;
Evans 1999b).

\acknowledgements

We thank the staff of the Owens Valley Millimeter array and the NICMOS GTO
team for their support both during and after the observations were
obtained, and the referee Sylvain Veilleux for many useful suggestions.
ASE also thanks M.  Shepherd, D. Frayer, and J. Surace for useful
discussion and assistance.  ASE was supported by NASA grant NAG 5-3042.
J.M.M. and D.-C.K. were supported by the Jet Propulsion Laboratory,
California Institute of Technology, under contract with NASA.

%\clearpage

%\vfill\eject
 
\centerline{Figure Captions}
 
\vskip 0.3in
 
\noindent
Figure 1. (a--c) {\it HST} NICMOS 1.1, 1.6, and 2.2 $\mu$m images of PKS
1345+12.  The images have peak intensities of 7.5, 11, and 29 $\mu$Jy for
the 1.1, 1.6, and 2.2 $\mu$m images, respectively. The speckle pattern
surrounding PKS 1345+12NW at 2.2 $\mu$m is a PSF artifact.  (d)
Continuum-subtracted CO($1\to0$) emission superimposed on the false-color
NICMOS image. The NICMOS data are displayed with blue as 1.1 $\mu$m, green
as 1.6 $\mu$m, and red as 2.2 $\mu$m. The CO data are plotted as 50\%,
60\%, 70\%, 80\%, 90\%, and 99\%, where 50\% corresponds to 3.9$\sigma$
rms and 100\% corresponds to a peak flux of 0.0339 Jy/beam. The CO
emission is unresolved, with a beam FWHM of 2.46$\times$1.95 at a 
position angle of -70.5$^{\rm o}$.

\vskip 0.1in
\noindent
Figure 2. Extracted CO($1\to0$) spectrum of PKS 1345+12.  The spectrum is
smoothed with a $\sim80$ km s$^{-1}$ filter and sampling of 40 km s$^{-1}$
intervals ($S_{\rm rms} \sim 0.005$ Jy).

\vfill\eject

\begin{deluxetable}{lll}
\tablenum{1}
\tablewidth{0pt}
\tablecaption{Peaks of Optical-to-Radio Emission in PKS 1345+12}
\tablehead{
\multicolumn{1}{c}{Source} &
\multicolumn{1}{c}{RA} &
\multicolumn{1}{c}{Dec}\nl
\multicolumn{1}{c}{} &
\multicolumn{2}{c}{(B1950.0)}\nl}
\startdata
13451+1232SE & 13:45:06.37 & 12:32:20.31\nl
13451+1232NW & 13:45:06.24 & 12:32:20.71\nl
2.3/8.4 GHz Radio$^1$ & 13:45:06.22 & 12:32:20.55 \nl
3mm/CO($1\to0$) & 13:45:06.21 & 12:32:20.31 \nl
\enddata
\tablenotetext{1}{The radio coordinates are taken from Ma (1998).}
\end{deluxetable}


\begin{thebibliography}{}

\bibitem[]{}
Armus, L., Heckman, T. M. \& Miley, G. H. 1987, \aj , 94, 831

\bibitem[]{}
Bryant, P. M. \& Scoville, N. Z., ApJ, in press

\bibitem[]{}
Evans, A. S. 1999a, Ap\&SS, in press (astro-ph/9903272)

\bibitem[]{} 
Evans, A. S. 1999b, in Highly Redshifted Radio Lines, ed.  C.  Carilli, S.
J. E. Radford, K. Menten, \& G. Langston (San Francisco: PASP), 156, 74

\bibitem[]{} 
Evans, A. S., Sanders, D. B., Mazzarella, J. M., \& Surace, J. A.
1999, ApJ, 511, 730

\bibitem[]{}
Fey, A. L., Clegg, A. W., \& Fomalont, E. B. 1996, ApJS, 105, 299

\bibitem[]{} 
de Grijp, M. H. K., Miley, G. K., \& Lub, J. 1987, A\&AS, 70,
95

\bibitem[]{}
Heckman, T. M. et al. 1986, ApJ, 311, 525

\bibitem[]{}
Joseph, R.D. \& Wright, G.S. 1985, MNRAS, 214, 87

\bibitem[]{}
Kim, D.-C. 1995, PhD Thesis, University of Hawaii at Manoa
 
\bibitem[]{}
Ma, C. 1998, AJ, 116, 516

\bibitem[]{}
MacKenty, J. W. \& Stockton, A. 1984, \apj, 283, 64

\bibitem[]{}
Mazzarella, J. M., Graham, J. R., Sanders, D. B., \& Djorgovski, S. 1993,
ApJ, 409, 170

\bibitem[]{} 
Mirabel, I. F., Sanders, D. B., \& Kaz$\grave e$s, I. 1989,
ApJ, 340, l9

\bibitem[]{} Murphy, T. W., Armus, L., Matthews, K., Soifer, B. T.,
Mazzarella, J. M.  \& Neugebauer, G. 1996, \aj , 111, 1025

\bibitem[]{}
Phillips, T. G. et al. 1987, ApJ, 322, L33

\bibitem[]{}
Radford, S. J. E., Solomon, P. M., \& Downes, D. 1991, ApJ, 368, L15
 
\bibitem[]{}
Rieke, M. 1999, in prep.

\bibitem[]{}
Sanders, D. B., Scoville, N. Z., \& Soifer, B. T. 1991, ApJ, 370, 158

\bibitem[]{}
Sanders, D.B., Soifer, B.T., Elias, J.H., Madore, B.F., Matthews, K.,
Neugebauer, G., \& Scoville, N.Z.  1988a, \apj , 325, 74

\bibitem[]{}
Sanders, D.B., Soifer, B.T., Elias, J.H., Neugebauer, G. \& Matthews, K.
1988b, \apj , 328, L35

\bibitem[]{}
Scoville, N. Z. et al. 1999, AJ, submitted

\bibitem[]{}
Scoville, N. Z., Carlstrom, J. C., Chandler, C. J., Phillips, J. A.,
Scott, S. L., Tilanus, R. P., \& Wang, Z. 1992, PASP, 105, 1482

\bibitem[]{}
Scoville, N. Z. \& Sanders, D. B. 1987, in Interstellar Processes, ed. D.
Hollenbach \& H. Thronson (Dordrecht: Reidel), 21

\bibitem[]{}
Scoville, N. Z., Sargent, A. I., Sanders, D. B., \& Soifer, B. T. 1991,
ApJ, 366, L5

\bibitem[]{}
Shepherd, M. C., Pearson, T. J., \& Taylor, G. B. 1995, BAAS, 27, 903 

\bibitem[]{}
Smith, E. P. \& Heckman, T. M. 1989a, ApJ, 341, 658
 
\bibitem[]{}
Smith, E. P. \& Heckman, T. M. 1989b ApJS, 69, 365
 
\bibitem[]{}
Solomon, P. M., Downes, D., Radford, S., \& Barrett, J. W. 1997,
ApJ, 478, 144

\bibitem[]{}
Steppe, H., Jeyakumar, S., Saikia, D. J. \& Salter, C. J. 1995,
A\&AS, 113, 409 

\bibitem[]{}
Stockton, A. \& MacKenty, J. W. 1983, Nature, 305, 678

\bibitem[]{}
Strong, A. W. et al. 1988, A\&A, 207, 1

\bibitem[]{}
Surace, J. A. \& Sanders, D. B. 1999, ApJ, 512, 162 

\bibitem[]{}
Surace, J. A., Sanders, D. B., Vacca, W. D., Veilleux, S., \& Mazzarella,
J. M. 1998, ApJ, 492, 116
 
\bibitem[]{}
Thompson, R. I, Rieke, M., Schneider, G., Hines, D. C., Corbin, M. R.
1998, ApJ, 492, L95

\bibitem[]{}
Veilleux, S., Kim, D.-C., Sanders, D. B., Mazzarella, J. M., \&
Soifer, B. T. 1995, ApJS, 98, 171

\bibitem[]{}
Veilleux, S. \& Sanders, D. B. 1999, in preparation

\bibitem[]{}
Veilleux, S., Sanders, D. B., \& Kim, D.-C. 1997, ApJ, 484, 92
 
\bibitem[]{}
Young, J. S. \& Scoville, N. Z. 1991, ARAA, 29, 581 

\end{thebibliography}
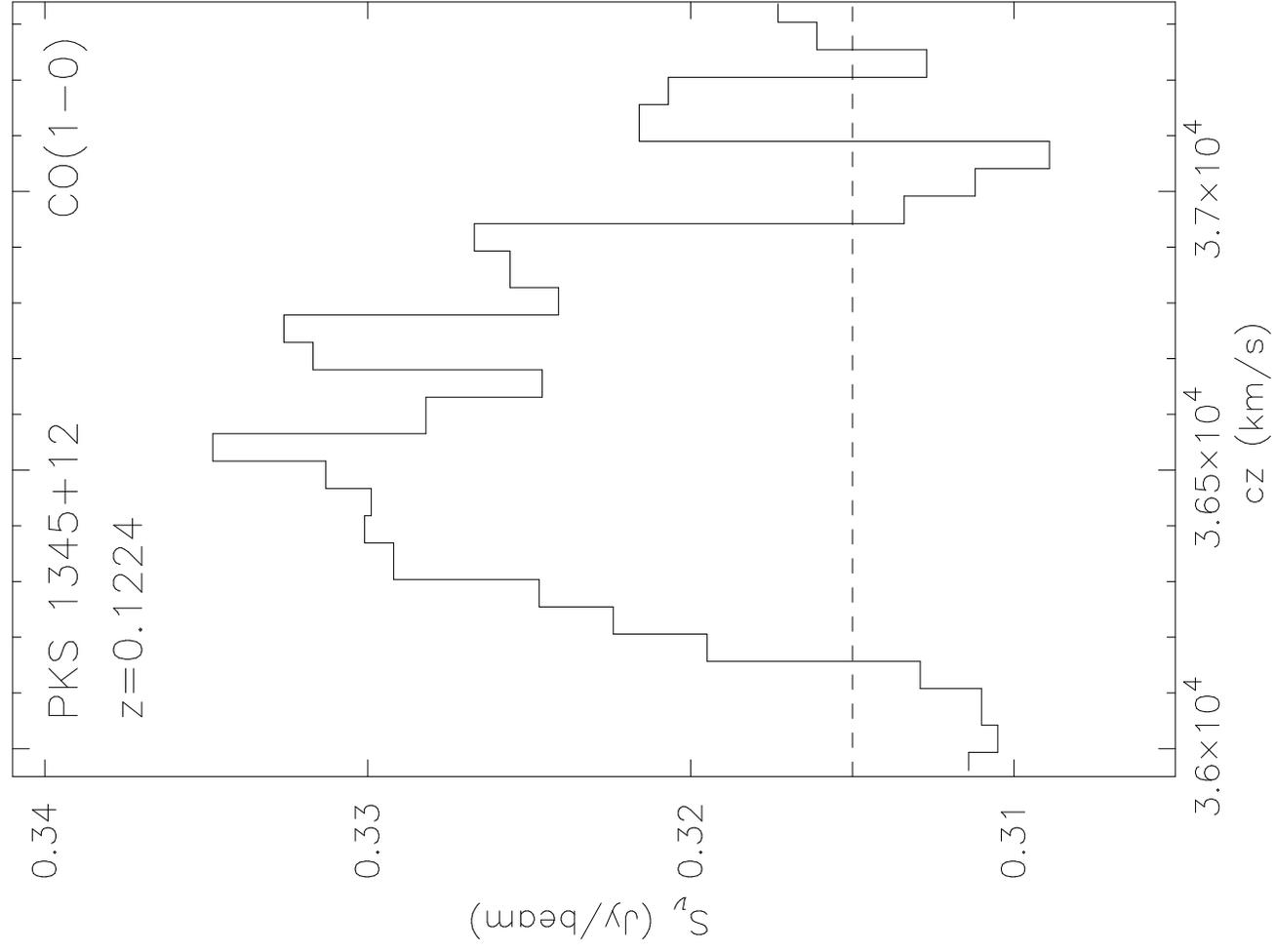
\end{document}